\documentstyle[12pt]{article}
\def\beq{\begin{equation}}
\def\eeq{\end{equation}}
\def\bea{\begin{eqnarray}}
\def\eea{\end{eqnarray}}
\def\bq{\begin{quote}}
\def\eq{\end{quote}}

\def\CQG{{\it Class. Quantum Gravity} }

\def\IJMP{{\it Int. J. Mod. Phys.} }

\def\NP{{\it Nucl. Phys.} }
\def\PL{{\it Phys. Lett.} }
\def\PR{{\it Phys. Rev.} }
\def\PRL{{\it Phys. Rev. Lett.} }

\def\gappeq{\mathrel{\rlap {\raise.5ex\hbox{$>$}}
{\lower.5ex\hbox{$\sim$}}}}

\def\lappeq{\mathrel{\rlap{\raise.5ex\hbox{$<$}}
{\lower.5ex\hbox{$\sim$}}}}

\begin{document}
\pagestyle{empty}

\begin{flushright}
DAMTP R-96/25  \\
DF/IST-3.96  \\
DM/IST-13/96 
\end{flushright}

\begin{center}
{\Large\bf Quantum Cosmological Multidimensional} \\
\vspace*{0.2cm}
{\Large\bf Einstein-Yang-Mills  Model} \\
\vspace*{0.2cm}
{\Large\bf  in a ${\bf R} \times S^3 \times S^d$ Topology}\\
\vspace*{0.7cm}
{\large\sf O. Bertolami}\footnote{e-mail: {\sf orfeu@cosmos.ist.utl.pt}}\\
Instituto Superior T\'ecnico \\Departamento de F\'\i sica \\
Av. Rovisco Pais, 1096 Lisboa Codex,
Portugal \\
\vspace*{0.5cm}
{\large\sf P.D. Fonseca}\footnote{e-mail: {\sf pdfonsec@math.ist.utl.pt}}\\
Instituto Superior T\'ecnico \\Departamento de Matem\'atica \\
Av. Rovisco Pais, 1096 Lisboa Codex,
Portugal \\
\vspace*{0.5cm}
and \\
\vspace*{0.5cm}
{\large\sf P.V. Moniz}\footnote{ e-mail: {\sf prlvm10@amtp.cam.ac.uk}}
\\
University of Cambridge, DAMTP \\
Silver Street, Cambridge, CB3 9EW, UK

\vspace*{0.7cm}
{\bf ABSTRACT} \\ \end{center}
\indent

The quantum cosmological version of the multidimensional
Einstein-Yang-Mills model in a ${\bf R} \times S^3 \times S^d$
topology is studied in the framework of the Hartle-Hawking
proposal. In contrast to previous work in the literature, 
we consider Yang-Mills field configurations with non-vanishing 
time-dependent components in both $S^3$ and $S^d$ spaces.
We obtain  stable compactifying solutions that do correspond to extrema 
of the Hartle-Hawking wave function of the Universe. 
Subsequently,
we also show that the regions where 4-dimensional metric behaves classically or 
quantum mechanically (i.e. regions where the metric is Lorentzian 
or Euclidean) will  depend on the number, $d$, of compact space dimensions.

\vfill\eject

\setcounter{page}{2}
\pagestyle{plain}

\vspace{0.5cm}

\section{Introduction}

\indent

The issue of compactification is central in multidimensional theories of
unification, such as generalized Kaluza -- Klein theories, Supergravity and 
Superstring theories. Consistency with known phenomenology requires 
that the extra
dimensions in these theories are Planck size and stable. A necessary condition
for the latter is the presence of matter with repulsive stresses to
counterbalance the collapsing thrust of gravity. For this purpose, magnetic
monopoles \cite{dss}, Casimir forces \cite{ac} and Yang-Mills fields
\cite{krt,bkm} have been considered. The situation with Yang-Mills fields is
particularly interesting as it illustrates well the importance of considering
non-vanishing  external-space components of the gauge fields, a point that 
has been disregarded in previous work in the literature. 
In fact, it was  shown in
Ref. \cite{bkm} that  it is precisely this feature that renders 
compactifying solutions
classically as well as semiclassicaly stable.

The main motivation for considering our study of compactification in
the context of quantum cosmology lies in ascertaining how this process
takes place. Indeed, this is crucial for extracting classical
predictions from any multidimensional unifying theories. In fact, no
cosmological description can be considered complete till specifying
the set of initial conditions for integration of the classical
equations of motion. Furthermore, since the quantum cosmological
approach of Hartle and Hawking \cite{hh} allows for a well defined
programme for establishing this set of initial conditions, it is quite
natural to extend this approach to the study of the issue of
compactification in higher-dimensional theories. This programme has
been already applied to many different quantum models of interest such
as massive scalar fields \cite{swh}, Yang-Mills fields \cite{bm},
massive vector fields \cite{bmon} as well as in supersymmetric models
(see Ref. \cite{pvm} for a review and a complete set of references)
and to the lowest order gravity-dilaton theory arising from string
theory \cite{bb}.  The generalization of the Hartle-Hawking programme
to higher spacetime dimensions has been considered previously for the
6-dimensional Einstein-Maxwell theory \cite{jh}, for gravity coupled
with a ($D-4$)th rank antisymmetric tensor field \cite{cw}, where the
stability of compactification was achieved thanks to the presence of a
magnetic monopole type configuration, and also to 11-dimensional
supergravity \cite{chinaprd}.

In this work a rather general and realistic setting to study the
compactification process is considered in the context of
Einstein-Yang-Mills multidimensional model of Ref. \cite{bkm} with an
$SO(N)$ gauge field in $D = 4 + d$ dimensions and an homogeneous and
(partially) isotropic spacetime with a ${\bf R} \times S^3 \times S^d$
topology. We aim to study the quantum mechanics of the coset
compactification of the $D$-dimensional spacetime ${\cal M}^D$
\begin{equation}
{\cal M}^D = {\bf R} \times G^{{\rm ext}}/H^{{\rm ext}} \times G^{{\rm int}}/
H^{{\rm int}}
\label{1}~,
\end{equation}
where $G^{\rm ext(int)} = SO(4) (SO(d+1))$ and $H^{{\rm ext(int)}} = 
SO(3) (SO(d))$ are
respectively the homogeneity and isotropy groups in 3($d$) dimensions. 
For this purpose we will seek compactifying solutions of the Wheeler-DeWitt 
equation for the Einstein-Yang-Mills cosmological model of Ref. \cite{bkm} 
in the framework of the Hartle-Hawking proposal.

In contrast to previous work in the literature \cite{jh,cw} we
consider Yang-Mills field configurations with {\em non-vanishing}
time-dependent components in {\em both} $S^3$ and $S^d$ spaces.  We
then derive an effective model by restricting the fields to be
homogeneous and isotropic. This construction will allow us to study in
detail the issue of compactification, which as discussed in
Ref. \cite{bkm}, depends crucially in the contribution of the {\em
external} gauge field components.  Our analysis of the resulting
Wheeler-DeWitt equation indicates that the regions where the metric is
Lorentzian or Euclidean do depend on the number, $d$, of internal
dimensions {\em and} on the potentials for the external and internal
components of the gauge field.  Furthermore, we show that stable
compactifying solutions do indeed correspond to the extrema of the
wave function of the Universe implying a correlation between
compactification of the extra dimensions and expansion of the
macroscopic spacetime. We should mention that an attractive feature of
our model is that it can be regarded as the bosonic sector of some
general unifying theories, implying that possibly most of the
conclusions of our quantum mechanical analysis of the compactification
process and of its stability will remain valid in those theories as
well.

This paper is then organized as follows.  In the next section we
present our Ans\"atze for the metric and for the gauge field (see
Refs. \cite{bkm, bmpv} for a general discussion) as well as the
resulting effective action which is the starting point of our
analysis. We also obtain in that section the Wheeler-DeWitt equation
of our effective model. In the Section 3 we present and discuss
compactifying solutions of the Wheeler-DeWitt equation and in Section
4 we discuss their interpretation. In Section 5 we present our
conclusions. We also include an Appendix where the mathematical
aspects of extending the Hartle-Hawking proposal to higher-dimensional
spacetimes is described, with emphasis in our model where
hypersurfaces are of $\Sigma_{D-1} \sim S^3 \times S^d$ type.

\vspace*{0.5cm}

\section{Effective Model and Wheeler-DeWitt Equation}

\indent 

We shall describe in this section our multidimensional
Einstein-Yang-Mills quantum cosmological model. Special emphasis will
be given to the differences between our model and others present in
the literature \cite{jh,cw,chinaprd}.  Namely, the gauge field in our
reduced model will have time-dependent spatial components on the
3-dimensional physical space. This contrasts with previous work on the
subject where either static magnetic monopole type configurations,
whose only non-vanishing components were the internal $d-$dimensional
ones \cite{jh,cw}, or scalar fields \cite{mel} were considered.  Our
approach provides therefore a somewhat more realistic model to study
the influence of higher dimensions on the evolution of the
4-dimensional physical spacetime. In addition, we shall also see how
{\em different} values for $d$, the number of internal space
dimensions, may induce fairly different physical situations.

Our model is derived from the generalized Kaluza-Klein action: 
\begin{equation}
S[\hat g_{\hat\mu\hat\nu}, \hat A_{\hat\mu}, \hat \chi] = 
S_{{\rm gr}}[\hat g_{\hat\mu\hat\nu}]
 + 
S_{{\rm gf}}[\hat g_{\hat\mu\hat\nu}, \hat A_{\hat\mu}]
 + 
S_{{\rm inf}}[\hat g_{\hat\mu\hat\nu},   \hat \chi]~,
\label{eq:2.1}
\end{equation}
with 
\begin{eqnarray}
S_{{\rm gr}}[\hat g_{\hat\mu\hat\nu}] & = & 
\frac{1}{16 \pi \hat k} \int_{{\cal M}^D} d \hat{x} 
\sqrt{-\hat g} (\hat R - 2 \hat \Lambda) ~, \label{eq:2.2a}
\\
S_{{\rm gf}}[\hat g_{\hat\mu\hat\nu}, \hat A_{\hat\mu}]
& = & 
\frac{1}{8\hat{e}^2} 
\int_{{\cal M}^D} d \hat{x} 
\sqrt{-\hat g} {\rm Tr} \hat F_{\hat\mu\hat\nu} 
\hat F^{\hat\mu\hat\nu} ~, \label{eq:2.3a} \\
S_{{\rm inf}}[\hat g_{\hat\mu\hat\nu},   \hat \chi]
& = & 
- \int_{{\cal M}^D} d \hat{x} 
\sqrt{-\hat g} \left[\frac{1}{2} 
\left(\partial_{\hat\mu} \hat \chi\right)^2 
+ \hat U \left(\hat \chi\right)
\right] ~, 
\label{eq:2.3b} 
\end{eqnarray}
where $\hat g$ is $\det 
\left(\hat g_{\hat\mu\hat\nu}\right)$, 
$\hat g_{\hat\mu\hat\nu}$ is the $D=4+d$ dimensional metric, 
$\hat R$, $\hat e$, $\hat k$ and $\hat \Lambda$ are, 
respectively, the scalar curvature, gauge coupling, 
 gravitational and cosmological constants 
in $D$ dimensions. In addition, the following 
filed variables are defined in ${\cal M}^D$: 
$\hat F_{\hat\mu\hat\nu} = 
\partial_{\hat\mu} \hat A_{\hat\nu} - 
\partial_{\hat\nu} \hat A_{\hat\mu} + 
\left[ \hat A_{\hat\mu}, \hat A_{\hat\nu}\right]$ 
is the field strenght and 
$\hat A_\mu$ denotes the components of the 
gauge field, $\hat \chi$ is the inflaton 
responsible for the inflationary expansion of the external 
space with $\hat U (\hat \chi)$ being  the potential  for $\hat \chi$. 
We assume that the potential $\hat U(\hat \chi)$ is bounded 
from below, that it has a global minimum and without loss of 
generality that $\hat U_{{\rm min}} = 0$. 
As first suggested in Refs. \cite{csl}, 
the splitting of the internal and external 
dimensions of space in the generalized Kaluza-Klein theory
(\ref{eq:2.1}) 
can have its origin in the
spontaneous symmetry 
breaking process, which is due to vacuum solutions corresponding to a
factorization of 
spacetime in a product of spaces. Assuming that is indeed the case, then:
\begin{equation}
 {\cal M}^D =  M^4 \times I^d,
\label{eq:1.1a}
\end{equation}
$M^4$ being the four-dimensional Minkowski  spacetime and 
$I^d$ a Planck-size $d-$dimensional compact space. For the cosmological 
setting we are interested in  consider instead 
\begin{equation}
{\cal M}^{4+d} = {\bf R} \times G^{{\rm ext}}/H^{{\rm ext}} \times 
G^{{\rm int}}/H^{{\rm int}}~,
\label{eq:1.2}
\end{equation}
admiting local coordinates $\hat x^{\hat\mu} =  
(t, x^i, \xi^m)$, where $\hat\mu = 0,1, \dots, 3+d$; 
$i=1,2,3$; $m=4, \ldots, d+3$; where {\bf R} denotes a timelike direction 
and $ G^{{\rm ext}}/H^{{\rm ext}}\left( 
G^{{\rm int}}/H^{{\rm int}}\right)$ the space of external 
(internal) spatial dimensions realized as a coset space of the external 
(internal) isometry group 
$ G^{{\rm ext }}\left( 
G^{{\rm int}}\right)$. 

We restrict ourselves to spatially homogeneous and (partially) 
isotropic field configurations, which means that these are symmetric under 
the action of the group 
$G^{{\rm ext }} \times  
G^{{\rm int}}$. Let the gauge group $\hat K$ of the $D-$dimensional 
theory be a simple Lie group. For definiteness, let us consider the case
with the gauge group $\hat K = SO(N), N \geq 3+d$ and 
\begin{equation}
{\cal M}^{4+d} = {\bf R} \times S^3 \times S^d,
\label{eq:2.4}
\end{equation}
where $S^3 (S^d)$ is the $3- (d-)$dimensional sphere. 
The group of spatial homogeneity and isotropy is, in this case: 
\begin{equation}
G^{{\rm HI}} = SO(4) \times SO(d+1)~,
\label{eq:2.5}
\end{equation}
while the group of spatial isotropy is 
\begin{equation}
H^{{\rm I}} = SO(3) \times SO(d),
\label{eq:2.6}
\end{equation}
which allows for the alternative realization of ${\cal M}^{4+d}$ 
\begin{eqnarray}
{\cal M}^{4+d} & 
= & {\bf R} \times SO(4)/SO(3) \times SO(d+1)/SO(d) \nonumber \\
& =& {\bf R} \times [SO(4)\times SO(d+1)]/[SO(3) \times SO(d)] ~.
\label{eq:2.7}
\end{eqnarray}

The field configurations associated with the above geometry were 
described in Ref. \cite{bkm}, using the theory of symmetric fields
(see also Refs. \cite{bm, bbmms, mms}). The most general 
form of a $SO(4) \times SO(d+1)-$invariant metric in $E^{4+d}$ as 
(\ref{eq:2.4}) reads 
\begin{equation}
\hat g = -\tilde N^2(t) dt^2 + \tilde a^2(t) 
\Sigma_{i=1}^3 \omega^i \omega^i 
+ b^2(t) \Sigma_{m=4}^{d+3} \omega^m \omega^m~,
\label{eq:2.11}
\end{equation}
where the scale factors $\tilde a(t), b(t)$ and the lapse function $\tilde N
(t)$ are 
arbitrary non-vanishing functions of time. Moreover,
$\omega^\alpha$ denote local moving coframes in $S^3 \times S^d$,   
$\Sigma_{i=1}^3 \omega^i \omega^i $ and 
$\Sigma_{m=4}^{d+3} \omega^m \omega^m$ 
coincide with the standard metrics $d \Omega_3^2$ and $d \Omega^2_d$ 
of  $3$ and $d-$dimensional spheres with 
local coordinates $(x^i, \xi^m)$, repectively. 

The $SO(4)\times SO(d+1)-$invariant ansatz for the 
inflaton field $\hat \chi$ reads
\begin{equation}
\hat \chi (t, x^i, \xi^m) = \hat \chi (t).
\label{eq:2.13}
\end{equation}
As for the $SO(4) \times SO(d+1)-${\em symmetric} gauge field, the following 
Ansatz is considered:
\begin{eqnarray}
 {\hat A} & = &  \frac{1}{2}\Sigma_{p,q=1}^{N-3-d}
B^{pq}(t) {\cal T}^{(N)}_{3+d+p\, 3+d+q}dt 
+ \frac{1}{2} \Sigma_{1 \leq i < j \leq 3} 
{\cal T}_{ij}^{(N)} \omega^{ij} \nonumber \\ 
& + &  \frac{1}{2} \Sigma_{4\leq m < n \leq 3}
  {\cal T}_{mn}^{(N)} \tilde \omega^{m-3\, n-3}
\nonumber \\
& + & 
\Sigma_{i=1}^3 \left[
 \frac{1}{4} f_0 (t) \Sigma_{j,k=1}^3  \epsilon_{jik}{\cal T}_{jk}^{(N)}
+ \frac{1}{2} 
\Sigma_{p=1}^{N-3-d} 
f_p (t)  {\cal T}^{(N)}_{i\, d+3+p}\right] \omega^i
\nonumber \\ 
&+ &  \Sigma_{m=4}^{d+3} 
\left[
\frac{1}{2} \Sigma_{q=1}^{N-3-d} g_q (t) 
 {\cal T}^{(N)}_{m\, d+3+q}\right] \omega^m~,
\label{eq:2.16}
\end{eqnarray}
where $f_0(t) f_p (t), p= 1,\ldots, N-3-d; g_q (t), q=1, \ldots, N-3-d; 
B^{pq}(t), 1 \leq p < q \leq N-3-d$ are arbitrary functions and 
${\cal T}_{pq}^{(N)}, 1 \leq p< q \leq N$ are the generators of the gauge 
group $SO(N)$. 
We have used the decomposition 
\begin{equation}
\omega = \Sigma_{\alpha = 1}^{d+3} \omega^{\alpha} T_{\alpha} +
\Sigma_{1 \leq i < j \leq 3} \omega^{ij} \frac{T_{ij}^{(4)}}{2} 
+ \Sigma_{1 \leq m < n \leq d} \tilde \omega^{mn}
\frac{\tilde T_{mn}^{(d+1)}}{2}  
\end{equation}
for the Cartan's one-form in $S^3\times S^d$. Here 
$T_{ij}^{(4)}, \tilde T_{mn}^{(d+1)}$ form a basis of the Lie algebra 
of $G^{{\rm HI}}$, $T_{\alpha} = \frac{T_{\alpha 4}^{(4)}}{2}, 
\alpha = 1,2,3$ and $T_{\alpha} = \frac{T_{\alpha - 3 \, d=1}^{(d+1)}}{2}, 
\alpha = 4, \ldots, d+3$.

Substituting the Ans\"atze (\ref{eq:2.11}), (\ref{eq:2.13}) and 
(\ref{eq:2.16}) into action 
(\ref{eq:2.1}) and performing the conformal changes 
\begin{eqnarray}
\tilde N^2(t) & = &  \left[\frac{b_0}{b(t)}\right]^d N^2 (t)~, 
\label{eq:2.12a} \\ 
\tilde a^2 (t) & = & 
\left[\frac{b_0}{b(t)}\right]^d a^2 (t)~, 
\label{eq:2.12b}
\end{eqnarray}
where $b_0$ denotes the equilibrium value of $b$, 
we obtain a one-dimensional effective reduced action for the functions of 
time that parametrize the {\em symmetric} field configurations \cite{bkm}:


\begin{eqnarray}
& S_{\rm eff} & [a, \psi, f_o, {\bf f}, {\bf g}, \chi, N, \hat B]  = 
16 \pi^2 \int dt N a^3 \left\{ -\frac{3}{8\pi k} 
\frac{1}{a^{2}} \left[\frac{\dot{a}}{N}\right]^2 + 
\frac{3}{32\pi k} \frac{1}{a^{2}} + \frac{1}{2} 
\left[
\frac{\dot{\psi}}{N}\right]^2  + 
 \frac{1}{2}\left[\frac{\dot{\chi}}{N}\right]^2 \right. \nonumber \\
&+ & \left.  e^{d\beta\psi} \frac{3}{4e^2}\frac{1}{a^{2}} 
\left( \frac{1}{2}\left[\frac{\dot{f_0}}{N}\right]^2 
+ \frac{1}{2}\left[\frac{ {\cal D}_t {\bf f}   }{N}\right]^2\right) 
+ e^{-2\beta\psi} \frac{d}{4e^2}\frac{1}{b_0^{2}} 
\frac{1}{2}\left[\frac{ {\cal D}_t {\bf g}   }{N}\right]^2 - 
W(a,\psi,f_0,{\bf f}, 
{\bf g}, \chi) \right\},
\label{eq:2.17}
\end{eqnarray}
with $k = \hat k /v_d b_0^d, e^2 = \hat e^2/v_d b_0^d, 
\beta = \sqrt{16 \pi k / d(d+2)}, v_d$ is the the volume 
of $S^d$ for $b=1$, and where we have used 
$\psi = \beta^{-1} \ln (b/b_0)$ and 
$\chi = \sqrt{v_d b_0^d} \hat \chi$ as the  dilaton\footnote{The scale 
factor $b(t)$  of the internal space induces a behaviour similar to the 
case of a minimally coupled scalar field. In fact, 
by introducing the field $\psi$ by $b \sim \exp \psi$, this 
quantity corresponds to the scalar field which appears in the 
harmonic expansion of the Kaluza-Klein theory.} and 
inflaton fields, respectively. In (\ref{eq:2.17}), the dots denote 
time derivatives and ${\cal D}_t$ is the covariant derivative with 
respect to the remnant $SO(N-3-d)$ gauge field $\hat B(t)$ in {\bf R}:
\begin{equation}
{\cal D}_t {\bf f}(t) = \frac{d}{d t} {\bf f(t)} + 
\hat B(t) {\bf f}(t), ~
{\cal D}_t {\bf g}(t) = \frac{d}{d t} {\bf g(t)} + 
\hat B(t) {\bf g}(t).
\label{eq:2.18}
\end{equation}
Notice that  $f_0(t), {\bf f} = \left\{ f_p \right\}$ represent the gauge 
field components in the 4-dimensional physical space-time, while 
${\bf g} = \left\{ g_q \right\}$  denotes the components in the 
space $I^d$ 
and $\hat B$ is an
$(N-3-d)\times (N-3-d)$ antisymmetric matrix 
$\hat B = (B_{pq})$. The potential $W$ in (\ref{eq:2.17}) is given by 
\begin{eqnarray}
W & = & e^{-d\beta\psi} \left[ -e^{-2\beta\psi}
\frac{1}{16\pi k}\frac{d(d-1)}{4}\frac{1}{b_0^{2}} + 
e^{-4\beta\psi} \frac{1}{b_0^{4}}\frac{d(d-1)}{8 e^2} 
V_2 ({\bf g})
+  \frac{\Lambda}{8\pi k} + U(\chi) 
\right] \nonumber \\ & +  & 
e^{-2\beta\psi} \frac{1}{(ab_0)^{2}} \frac{3d}{32 e^2} 
({\bf f}\cdot 
{\bf g})^2 + e^{d\beta\psi} \frac{3}{4 e^2 a^4} 
V_1(f_0, {\bf f})~,
\label{eq:2.19}
\end{eqnarray}
where 
$\Lambda = v_d b_0^d \hat \Lambda$, $U(\chi) = v_d b_0^d 
\hat U \left(\hat \chi / \sqrt{v_d b^d_0}\right)$ and 
\begin{eqnarray}
V_1 (f_0, {\bf f}) & =  & \frac{1}{8} 
\left[ \left( f_0^2 + {\bf f}^2 - 1 \right)^2 + 
4 f_0^2 {\bf f}^2 \right],
\label{eq:2.20a} \\
V_2 ({\bf g}) & =  & \frac{1}{8} \left( {\bf g}^2  - 1\right)^2~,
\label{eq:2.20b}
\end{eqnarray}
are related with the external and internal components of the gauge field, 
respectively.
Variables $N$ and $\hat B$ are Lagrange multipliers associated 
with the symmetries of the effective action (\ref{eq:2.17}). 
The lapse function $N$ is associated 
with the  invariance of $S_{{\rm eff}}$ under arbitrary time 
reparametrizations, while $\hat B$ is connected with the 
local remnant $SO(N-d-3)$ gauge invariance. The equations of motion 
for the physical variables $a, \psi, \chi, f_0, {\bf f}, {\bf g}$ 
can be found in Ref. \cite{bkm}.

The canonical conjugate momenta associated with the canonical
variables in model (\ref{eq:2.17}) are given by:  
\begin{eqnarray}
\pi_a & = & - \frac{12\pi}{k}\frac{a}{N} \dot{a},~
\pi_{\psi} = 16 \pi^2 \frac{a^3}{N}\dot{\psi},~
\pi_{\chi} = 16 \pi^2 \frac{a^3}{N}\dot{\chi},~
\label{eq:1.7} \\
\pi_{f_0} & = & \frac{12 \pi^2}{e^2} e^{d\beta\psi} 
\frac{a}{N} \dot{f}_0,~
{\bf{\pi_f}}  =  \frac{12 \pi^2}{e^2} e^{d\beta\psi} 
\frac{a}{N} {\cal D}_t {\bf f},~
{\bf \pi_g}  =  \frac{4\pi^2}{e^2 b_0^2} e^{-2\beta\psi} 
\frac{a^3}{N} {\cal D}_t {\bf g}~.
\label{eq:1.8}
\end{eqnarray}
For simplicity we replace\footnote{The replacement $\psi \rightarrow 
\phi$ and $\chi \rightarrow \xi$ is a mere rescaling, 
while introducing $\mu \rightarrow \ln a$ for the scale factor can 
bring some advantages. In fact, 
the minisuperaspace metric becomes then proportional to 
${\rm diag} (1, -1)$ with useful consequences as far as the 
Wheeler-DeWitt equation is concerned \cite{jhjesu}.}
the variables $(a, \psi, \chi)$ by  the new variables $(\mu, \phi, \xi)$: 
\begin{equation}
a = e^\mu \left(\frac{k}{6\pi}\right)^{\frac{1}{2}}~,
\psi = \phi \left(\frac{3}{4\pi k }\right)^{\frac{1}{2}}~,
\chi = \xi \left(\frac{3}{4\pi k }\right)^{\frac{1}{2}}~.
\label{eq:1.9}
\end{equation}
The corresponding new conjugate momenta then read
\begin{equation}
\pi_\mu = - \left(\frac{2k}{3 \pi}\right)^{\frac{1}{2}} 
\frac{e^{3\mu}}{N} \dot{\mu}~,
\pi_\phi = \left(\frac{2k}{3 \pi}\right)^{\frac{1}{2}} 
\frac{e^{3\mu}}{N} \dot{\phi}~,
\pi_\xi = \left(\frac{2k}{3 \pi}\right)^{\frac{1}{2}} 
\frac{e^{3\mu}}{N} \dot{\xi}~.
\label{eq:1.10}
\end{equation}
The Hamiltonian and $SO(N-3-d)$ gauge 
constraints are then obtained by varying (\ref{eq:2.17}) 
with respect to $N$ and $\hat B$, and  
in terms of the momenta (\ref{eq:1.10}) are given by:
\begin{equation}
- \pi_\mu^2 -  e^{4\mu} + \pi_\phi^2 + \pi^2_\xi + 
 e^{2\mu-d\alpha\phi} \frac{e^2}{6\pi^2}
\left[\pi_{f_0}^2  + {\bf \pi_f}^2\right]
+ e^{2\alpha\phi} 
\frac{3 e^2 b_0^2}{d \pi k} 
{\bf \pi_g}^2 +  e^{6\mu} \left(\frac{4 k}{3}\right)^2W = 0~,
\label{eq:1.13a}
\end{equation}
\begin{equation}
\pi_{f_p} f_q + \pi_{g_p} g_q 
- \pi_{f_q} f_p - \pi_{g_q} g_p = 0~,
\label{eq:1.13b}
\end{equation}
where $\alpha = \sqrt{12/d (d+2)}$. 

The canonical quantization follows by promoting  
the conjugate momenta into operators as 
\begin{equation}
\pi_\mu \mapsto -i\frac{\partial}{\partial \, \mu}, ~
\pi_\phi \mapsto -i\frac{\partial}{\partial \, \phi}, ~
\pi_\xi \mapsto -i\frac{\partial}{\partial \, \xi}, ~
\pi_{f_0} \mapsto -i\frac{\partial}{\partial \, f_0}, ~
{\bf \pi_f} \mapsto -i\frac{\partial}{\partial \, {\bf f}}, ~
{\bf \pi_g} \mapsto -i\frac{\partial}{\partial \, {\bf g}} ~.
\label{eq:1.15}
\end{equation}
The Hamiltonian constraint (\ref{eq:1.13a}) is then 
quantized to yield the Wheeler-DeWitt equation: 
\begin{equation}
\left\{\frac{\partial^2}{\partial \mu^2} 
-  e^{4\mu} - \frac{\partial^2}{\partial \phi^2} -  
\frac{\partial^2}{\partial \xi^2} -  
 e^{2\mu-d\alpha\phi} \frac{e^2}{6\pi^2}
\left[\frac{\partial^2}{\partial f_0^2} + 
\frac{\partial^2}{\partial {\bf f}^2} \right]
- e^{2\alpha\phi} 
\frac{3 e^2 b_0^2}{d \pi k} 
\frac{\partial^2}{\partial {\bf g}^2} +  
e^{6\mu} \left(\frac{4 k}{3}\right)^2 W\right\} \Psi = 0~,
\label{eq:1.16}
\end{equation}
where in the usual parametrization of the factor ordering ambiguity, 
$\pi_\mu^2 \mapsto - \mu^{-p} \frac{\partial}{\partial \mu} 
\left(\mu^p  \frac{\partial}{\partial \mu} \right)$, we have set $p=0$.

The richness of the effective model (\ref{eq:2.17}) and the
corresponding Wheeler-DeWitt equation (\ref{eq:1.16}) is quite
evident. In this reduced model the gauge field has non-vanishing
time-dependent components in {\em both} the external and internal
spaces. Moreover, we have also two time-dependent scalar fields, the
dilaton and the inflaton. This contrasts with previous work in the
literature, where either static magnetic monopole configurations with
non-zero components only in $I^d$ or scalar fields were present.  Our
model allows thus to consider several possibilities.

Aiming to study the compactification process we shall focus our
analysis on the variables $\mu$ and $\phi$ and the contributions to
the potential $W$ from the gauge field. This choice is justifiable as
it can be seen from (\ref{eq:1.16}) that the kinetic term for the
external components of the gauge field is suppressed in an expanding
Universe, while for the internal components the kinetic term is not
relevant as compactifying solutions require ${\bf g}$ to seat at the
extremum of the potential $V_2({\bf g})$ \cite{bkm}. In doing that, we
shall keep the inflaton field frozen as it has been shown that this
field does not affect the compactification process \cite{bkm}. Of
course, we could instead consider taking $\mu$ and $\chi$ as the
physically relevant variables and freeze the remaining ones and
actually models of this type have been studied in Ref. \cite{mel}.

Hence, in what follows we shall restrict ourselves instead to the
study of compactification and hence concentrate our study on the
sub-system where the relevant variables are $\mu$ and $\phi$.  Hence,
it requires solving the Wheeler-DeWitt equation (\ref{eq:1.16}) for
static vacuum configuration of the gauge and inflaton fields:
\begin{equation}
\xi = \xi^v, ~f_0 = f_0^v, ~{\bf f} = {\bf f}^v, ~{\bf g} = {\bf g}^v = 
{\bf 0}~;
\label{eq:1.17}
\end{equation}
we also assume that $U(\xi^v) = 0$ and that {\bf f} and {\bf g} are 
orthogonal. The notation 
$v_1 \equiv V_1 (f_0^v, {\bf f}^v)$ and 
$v_2 \equiv V_2 ({\bf g}^v)=\frac{1}{8}$ will be used throughout
this paper. 
The Wheeler-DeWitt equation suitable for the study of 
compactification is the following: 
\begin{equation}
\left[ \frac{\partial^2}{\partial \mu^2} 
-  \frac{\partial^2}{\partial \phi^2} + U(\mu, \phi) \right] \Psi(\mu,\phi) 
= 0,
\label{eq:1.18a}
\end{equation}
where
\begin{equation}
U(\mu, \phi) = e^{6\mu}  \left(\frac{4 k}{3}\right)^2 
\Omega (\mu, \phi) - e^{4 \mu},
\label{eq:1.18b}
\end{equation}
and 
\begin{equation}
\Omega (\mu, \phi) = 
e^{-d\alpha\phi} \left[ -e^{-2\alpha\phi}
\frac{1}{16\pi k}\frac{d(d-1)}{4}\frac{1}{b_0^{2}} + 
e^{-4\alpha\phi} \frac{1}{b_0^{4}}\frac{d(d-1)}{8 e^2} 
v_2 
+  \frac{\Lambda}{8\pi k}  
\right] + 
  e^{d\alpha\phi -4 \mu} \left(\frac{6\pi}{k}\right)^2 
\frac{3}{4 e^2 } v_1 ~.
\label{eq:1.18c}
\end{equation}

The scenario associated with this choice is analogous to the ones of
Refs. \cite{jh,cw,chinaprd,chie}, with the novel feature of taking
into account the external components of the gauge field.  As it will
be seen, the last term in (\ref{eq:1.18c}) is central in our model and
constitutes one of the {\em major differences} with respect to, for
instance, Ref. \cite{jh}. Indeed, it is precisely this term sets the
dependence of early Universe scenarios ($\mu \ll 0$, i.e. $a
\rightarrow 0$) on different values of $d$ and $v_1$, brought about by
the gauge field components in the 4-dimensional spacetime.

Moreover, as it will be discussed in the next section, it is the term
$e^{d\alpha\phi -4\mu} \left(\frac{6\pi}{k}\right)^2
\frac{3}{4e^2}v_1$ in (\ref{eq:1.18c}) that establishes that the
external spatial dimensions and the internal $d$-dimensions are at the
{\em same footing} in the early Universe prior to compactification,
i.e., when $\mu \ll 0$. It is only through the expansion of the
external dimensions (increase of $\mu$) that compactification ($b
\rightarrow b_0$) is achieved. Thus, it is the dynamics of the
$3-$dimensional physical space which induces the evolution of $I^d$
towards compactification.  Furthermore, we shall see how different
values for $v_1$ and $d$ do lead to different quantum scenarios,
i.e. solutions of the Wheeler-DeWitt equation, whose physical features
can be compared with those of Refs. \cite{jh,cw}.

\section{Solutions with dynamical compactification}

\indent

In this section we shall establish the boundary conditions for the
Wheeler-DeWitt equation (\ref{eq:1.18a})-(\ref{eq:1.18c}) and obtain
solutions with dynamical compactification for certain regions of the
$\mu \phi$-plane.  Let us first address the latter issue, i.e., the
scenario for dynamical compactification in our model.

As discussed in Ref. \cite{bkm}, from the classical point of view, different 
values for the cosmological constant $\Lambda$ lead to different compactifying
scenarios. Indeed, for $\Lambda>c_2/16\pi k$  
($c_2=[(d+2)^2(d-1)/(d+4)]e^2/16v_2$) there are no compactifying solutions
and for
\beq
\frac{c_1}{16\pi k}<\Lambda<\frac{c_2}{16\pi k}
\eeq
($c_1=d(d-1)e^2/16v_2$) a compactifying solution exists which is classicaly
stable, but semiclassically unstable. Finally, a value of 
$\Lambda<c_1/16\pi k$ implies that the value of the effective 
4-dimensional cosmological constant, 
$\Lambda^{(4)}= 8\pi k \Omega(\infty,\phi)$, is negative (see Figure 1).
Since the 4-dimensional cosmological constant, $\Lambda^{(4)}$, must satisfy 
the bound
\beq
|\Lambda^{(4)}|<10^{-120}\frac{1}{16\pi k}~,
\eeq
we are led to choose $\Lambda=c_1/16\pi k$. On the other hand, since we 
are interested in compactifying solutions, for which $\phi\approx 0$, 
we shall take 
$\Lambda$ such that $\phi=0$ corresponds to the absolute minimum of 
(\ref{eq:1.18b}). This corresponds to $b_0^2 = \frac{16\pi k v_2}{e^2}$, and 
the fine-tuning \cite{bkm}
\beq
\Lambda=\frac{d(d-1)}{16b_0^2}~.
\label{eq:2.2}
\eeq
The potential (\ref{eq:1.18b}) simplifies then to
\beq
U(\mu,\phi)=e^{6 \mu-d \alpha \phi} \frac{2k \Lambda}{9 \pi}\left( e^{-2\alpha
\phi}-1 \right) ^2-e^{4\mu}+e^{2\mu+d \alpha
\phi}\frac{3\pi}{k}\frac{v_1}{v_2}b_0^2 ~,
\label{eq:3.3}
\eeq
and its form is shown in Figure 2. Moreover,
as can be seen from the plot of $\Omega(\mu=$constant$,\phi)$ in Figure 3 (cf.
eq. (\ref{eq:1.18b})),
for $\mu$ greater than a critical value, $\mu_{\rm c}$, the potential 
$U(\mu,\phi)$
has a local maximum, $\phi_{\rm max}$, given approximately by 
$e^{-2 \alpha \phi_{\rm max}}=d/(d+4)$.
This critical value arises from the last term in (\ref{eq:3.3}) and in first 
order approximation is given by 
\beq
a_c^4 \sim e^{4 \mu_{\rm c}}=\frac{-B-\sqrt{B^2-4AC}}{2A}~,
\eeq
where
\beq
\begin{array}{l}\displaystyle
A=-\alpha A_1 A_2 \left(\frac{2k\Lambda}{9\pi}\right)^2 \\ \displaystyle
B= d \frac{2\Lambda}{3}\frac{v_1}{v_2}b_0^2
\left[dA_1e^{2d\alpha \phi_0}\left(\frac{1}{\phi_{\rm max}-\phi_0}
+\alpha d\right)+
A_2e^{2d\alpha \phi_{\rm max}}\left(\frac{1}{\phi_{\rm max}-\phi_0}
-\alpha d\right)\right]
\\ \displaystyle
C=\alpha d^5 e^{2d\alpha \phi_0}
\left(\frac{3\pi}{k}\frac{v_1}{v_2}b_0^2\right)^2,
\end{array}
\eeq
with $A_1=-8e^{-2 \alpha \phi_{\rm max}}=\frac{-8d}{d+4}$, 
$A_2=[(d+4)^2e^{-4\alpha \phi_0}-d^2]/2$ and 
$e^{-2\alpha \phi_0}=[(d+2)^2-\break \sqrt{(d+2)^4-d^2(d+4)^2}]/(d+4)^2$.

We now turn to the discussion of the boundary conditions for the Wheeler-DeWitt
equation \footnote{A discussion on the mathematical aspects of
generalizing the Hartle-Hawking no-boundary proposal to higher
dimensions can be found in the Appendix.}.  We shall use the path
integral representation for the ground-state of the Universe \beq
\Psi[\mu,\phi]=\int_{C}{D\mu D\phi \exp(-S_{\rm E})}~,
\label{eq:3.4}
\eeq
which does allow us to evaluate $\Psi(\mu,\phi)$ close to
$\mu=-\infty$.  In here, $S_{\rm E}=-iS_{\rm eff}$ is the Euclidean
action, obtained throught the effective action (\ref{eq:2.17}) and
taking $d\tau=iNdt$

\beq
S_E=\int{d\tau \frac{6\pi}{k} \left[
-a \dot{a}^2 +a^3\dot{\phi}^2-\frac{a}{4}+
a^3e^{-d\alpha \phi}(e^{-2\alpha\phi}-1)^2\frac{\Lambda}{3}
+\frac{1}{a}e^{d\alpha\phi}
\frac{2\pi k}{e^2}v_1
\right]}.
\label{eq:euclidean action}
\eeq 
To ensure that the sum $C$ does corresponds to compact 
$(d+4)$-metrics we must impose conditions on $\tilde{a}(t)$ and $b(t)$ at
$\tau=0$ (where $\tau$ is the euclidean time $d\tau=iNdt$),
such that the Euclidean metric
\beq
\hat{g}=d\tau^2+\tilde{a}^2(\tau)\Sigma_{i=1}^3 \omega^i \omega^i 
+ b^2(\tau) \Sigma_{m=4}^{d+3}\omega^m \omega^m~,
\label{eq:3.6}
\eeq is compact. In Ref. \cite{jh} the following conditions were
suggested: $\tilde{a}=0$, $b>0$, $\frac{d\tilde{a}}{d\tau}=1$ and
$\frac{db}{d\tau}=0$ at $\tau=0$, which can also be inferred from the
regularity of the Euclidean equations of motion \cite{cw}.  Notice
that physical reasons, such as the vanishing of the internal gauge
field components and of the gravitational coupling in 4-dimensions,
prevent the interchange of these conditions. Clearly, this approach to
select the boundary conditions to the Hartle-Hawking wave function is
not quite correct from the quantum point of view as it implies a
simultaneous fixing of both canonical and corresponding conjugated
momentum variables.

As far as our reduced model (see eq. (\ref{eq:2.17})) 
is concerned, consistent boundary conditions
can be implemented as follows. Let us first 
 point out that our reduced model is similar to a closed
Friedmann-Robertson-Walker model with a scalar field $\psi$ (or
$\phi$) \cite{jh}. Hence,  our   boundary conditions 
 which are consistent with a 4-geometry closing off in a regular way and with 
regular field configurations: $a(0) = 0$ and 
$\frac{d \psi}{d \tau} (0) = 0$. The next step 
is to note that
the  corresponding 
constraint (Friedmann) equation in our model 
imples that $ \frac{d a}{d \tau}(0) = 1$ \cite{g1,g2}, i.e. 
the condition $a(0) =0$ is equivalent to $ \frac{d a}{d \tau}(0) = 1$. 
In addition, \cite{swh,jhjesu} $\frac{d \psi}{d \tau} (0) = 0$
 leads, using $ \psi \sim \ln b$, to 
$\frac{d b}{d \tau}(0) = 0$ and   $b(0)>0$. 
It is important to realize that the geometries summed over in the 
path integral 
will be closed at $\tau =0$ for the 4-dimensional physical spacetime, 
but  generally not regular, and also that  the geometries will be 
regular at $\tau = 0$ for the extra $d$-dimensional space. 
>From the constraint equation, the other condition 
$\frac{d a}{d \tau} (0) = 1$ (regularity) 
will hold at saddle points, and similarly for $b(0) >0$ which 
will follow from the corresponding regularity of the equations of motion 
\cite{cw}.

Thus, integrating (\ref{eq:euclidean action}) from an initial 
point $\tau=0$ to $\Delta\tau$, a very close point to $\tau=0$, we get
\beq
S_E=\int_0 ^{\Delta\tau}{d\tau \frac{6\pi}{k} \left[
-\tau -\frac{\tau}{4}+
\tau^3e^{-d\alpha \phi}(e^{-2\alpha\phi}-1)^2\frac{\Lambda}{3}
+\frac{1}{\tau}e^{d\alpha\phi}\frac{2\pi k}{e^2}v_1
\right]},
\eeq
where we used $a\approx \tau$ close to $\tau=0$. Finally, by
setting $a=e^{\mu}\sqrt{k/6\pi}$, the integration yields\\
\beq
S_{\rm E}=\left\{\begin{array}{cl}
\displaystyle -\frac{5}{8}e^{2\mu}+e^{4\mu-d\alpha \phi}
\left(e^{-2\alpha \phi}-1\right)^2\frac{k\Lambda}{72\pi}
&\mbox{, for $v_1=0$}~,
\\
+\infty &\mbox{, for $v_1\neq 0$}~.
\end{array}\right.
\label{eq:3.9}
\eeq

Since, with a suitable choice of the metric, we can have
 $\Psi=e^{-S_{\rm E}}$ near the past null infinity 
(see ref. \cite{jh}), $\Im^-$, we can 
easily obtain the boundary conditions. This analysis is simplified 
by introducing the following new variables:
\beq
\begin{array}{l}
x=e^{\mu}\sinh\phi \\
y=e^{\mu}\cosh\phi~,
\end{array}
\eeq 
such that the past null infinity $\Im^-$ now corresponds to the
lines $x=y$ and $x=-y$. The boundary conditions on $\Im^-$, that are
shown in Table 1, can be easily obtained from (\ref{eq:3.9}).  For all
over $\Im^-$, the normal derivative vanishes, $\frac{\partial
\Psi}{\partial n}=0$.

\[
\begin{tabular}{|l|l|r|}   \hline\hline
   {  }                 &{$v_1=0$}           & {$v_1\neq 0$} \\ \hline
on $\Im^-$ and $\phi<0$ &  $\Psi=0 (1)$ for $d<19$ ($d\ge 19$)  & $\Psi=0$  \\
on $\Im^-$ and $\phi>0$ &  $\Psi=1$                            & $\Psi=0$  \\
\hline\hline
\end{tabular}
\]
\vspace{0.2 cm}
\centerline{{\bf Table 1}: Boundary conditions on $\Im^-$ for $\Psi$.}
\vspace{0.2 cm}

Let us now further proceed with our search for solutions to the
Wheeler-DeWitt equation. In this situation, one must generally begin
by determining the regions where the solution is oscillatory and where
it is exponential. This can be heuristically done by examining the
regions where for surfaces of constant $U$, the minisuperspace metric
$ds^2=d\mu^2 - d\phi^2$ is either spacelike ($ds^2 > 0$) or timelike
($ds^2 < 0$):

In spacelike regions we can locally perform a Lorentz-type 
transformation to new coordinates $(\tilde\mu,\tilde\phi)$:
\beq
\begin{array}{l}
\tilde{\mu}=\mu\cosh\theta-\phi\sinh\theta\\
\tilde{\phi}=-\mu\sinh\theta+\phi\cosh\theta~,
\label{eq:4.7}
\end{array}
\eeq
where $\theta$ is a constant,
such that the 
surfaces of constant $U$ are parallel to the $\tilde\phi$ axis. The potential 
will then depend, at least locally, only on $\tilde\mu$  and the 
Wheeler-DeWitt equation can be rewritten as 
\beq
\left[\frac{\partial^2}{\partial\tilde{\mu}^2}-\frac{\partial^2}{\partial\tilde
{\phi}^2}+U(\tilde{\mu})\right] \Psi(\tilde\mu, \tilde \phi) = 0~,
\eeq
and $\Psi$ will be oscillatory if $U>0$ and exponential type if $U<0$, 
assuming that its dependence on $\tilde{\phi}$ is small.

Similarly, when the surfaces of constant $U$ correpond to
timelike regions of the minisuperspace metric, a Lorent-type
transformation can rotate coordinates $(\mu,\phi)$ such that they become 
parallel to the $\tilde{\mu}$ axis. The potential, $U$, will then depend only 
on $\tilde{\phi}$, and $\Psi$ will be exponential type for $U<0$ and 
oscillatory type for $U>0$, assuming now that the wave function dependence

on $\tilde{\mu}$ is small.
The surfaces $U=0$ depend on the relation $\frac{v_1}{v_2}$ and are given by
 the expression
\beq
e^{2\mu}=\frac{9\pi}{4k\Lambda}\frac{e^{d\alpha\phi}}
{\left(e^{-2\alpha\phi}-1\right)^2}
\left(1\pm\left[1-\frac{d(d-1)}{6}\frac{v_1}{v_2}
\left(e^{-2\alpha\phi}-1\right)^2\right]^{1/2}\right)~.
\eeq

These surfaces (see Figure 4) provide all points for which a
Euclidean solution can be smoothly matched into a Lorentzian one,
that is $\dot{\mu}=\dot{\phi}$ (the extrinsic curvature being continuous).
For $\frac{v_1}{v_2}=0$ we recover the result found in 
Ref. \cite{jh}.
In order to further charactherize the regions where solutions are oscillatory
or 
exponential, we further summarize 
the asymptotic branches of the surface $U=0$  as follows:

\begin{description}

\item[{\it (i)}] For $v_1/v_2=0$ and $\phi\rightarrow +\infty$,
$b\rightarrow +\infty$, we have $e^{2\mu}\rightarrow
\frac{9\pi}{2k\Lambda} e^{d\alpha\phi},
\tilde{a}\rightarrow\sqrt{\frac{3}{4\Lambda}}$.

\item[{\it (ii)}] When $\phi\rightarrow -\infty$, $b\rightarrow 0$, we
have $e^{2\mu}\rightarrow \frac{9\pi}{2k\Lambda}e^{\alpha(d+4)\phi}$,
$\tilde{a}\rightarrow 0$.

\item[{\it (iii)}] Finally, when $\phi\rightarrow 0$, $b\rightarrow
b_0$, we obtain $e^{2\mu}\propto \phi^{-1}$, $\tilde{a}\rightarrow 0$.

\item[{\it (iv)}] For $v_1/v_2\neq 0$ only the asymptotic branch
$\phi\rightarrow 0$ survives.

\end{description}

However, 
besides the surfaces of constant $U$ that correspond to timelike or spacelike
regions, we have also
to look for the curves of constant $U$ surfaces for which the
minisuperspace metric is null,  $\frac{d\mu}{d\phi}=\pm1$.
The expression for these curves is given by $\frac{\partial U}{\partial\mu}=
\pm\frac{\partial U}{\partial\phi}$, that is
\beq \displaystyle
e^{2\mu}=\frac{9\pi}{k\Lambda}e^{d\alpha\phi}
{1\mbox{$``\pm$''}\left[1-\frac{d(d-1)}{96}\frac{v_1}{v_2}
\left(e^{-2\alpha\phi}-1\right)
(2\mp d\alpha)\left[e^{-2\alpha\phi}\left(6\pm\alpha(d+4)\right)-
(6\pm d\alpha)\right]\right]^{1/2}\over
\left(e^{-2\alpha\phi}-1\right)
\left[e^{-2\alpha\phi}\left(6\pm\alpha(d+4)\right)-(6\pm d\alpha)\right]}~,
\label{eq:3.10a}
\eeq
where the sign $``\pm$'' is independent of the remaining ones
 appearing in (\ref{eq:3.10a}).
It is quite important  to point out that the sign of one of the terms 
in (\ref{eq:3.10a}) {\em depends} on the number of 
extra dimensions, $d$:
\beq
\begin{array}{cl}
6-\alpha(d+4)>0 &\mbox{for $d\ge 4$}\\
6-\alpha(d+4)<0 &\mbox{for $d< 4$}~.
\end{array}
\eeq
This implies that there will be different solutions for different
values of $d$.
As far as the asymptotic branches of (\ref{eq:3.10a}) are concerned, we have
the following: 

\begin{description}

\item[{\it (i)}] For $\phi\rightarrow +\infty$, $b\rightarrow\infty$,
we have the asymptotic branch
$e^{2\mu}\rightarrow\frac{9\pi}{k\Lambda}
e^{d\alpha\phi}(\frac{1+\sqrt{C_+}}{6-\alpha d})$,
$\tilde{a}\propto\Lambda^{-1/2}$, where $C_{\pm}=1-\frac{d(d-1)}{96}
\frac{v_1}{v_2}(2\mp d\alpha)(6\pm d\alpha)$.

\item[{\it (ii)}] If $v_1/v_2$ verifies the condition
$v_1/v_2<96/[d(d-1) (2+d\alpha)(6-d\alpha)]$, then there are two other
asymptotic branches:
$e^{2\mu}\rightarrow\frac{9\pi}{k\Lambda}e^{d\alpha\phi}
(\frac{1\pm\sqrt{C_-}}{6-d\alpha})$, $\tilde{a}\propto\Lambda^{-1/2}$.

\item[{\it (iii)}] For $\phi\rightarrow -\infty$, $b\rightarrow 0$,
and $v_1/v_2=0$ we have
$e^{2\mu}\rightarrow\frac{9\pi}{k\Lambda}e^{(d+4)\alpha\phi}
\frac{2}{6\pm\alpha(d+4)}$, $\tilde{a}\rightarrow 0$. The lower sign
branch exists only for $d\ge4$.

\item[{\it (iv)}] When $\phi\rightarrow -\infty$ and
$v_1/v_2\neq0$, we have
$e^{2\mu}\rightarrow\frac{9\pi}{k\Lambda}e^{(d+2)\alpha\phi}
\sqrt{\frac{d(d-1)}{96}\frac{(-2\pm d\alpha)}{6\pm\alpha(d+4)}
\frac{v_1}{v_2}}$, $\tilde{a}\rightarrow 0$, and the lower sign branch
exists only for $d<4$.

\item[{\it (v)}] Finally, when $\phi\rightarrow 0$, $b\rightarrow
b_0$, then $e^{2\mu}\propto \phi^{-1}$.

\item[{\it (vi)}] There is an additional asymptotic branch for
$\phi\rightarrow\phi_{\pm}$, where $\exp(-2\alpha\phi_{\pm})=(6\pm
d\alpha)/[6\pm\alpha(d+4)]$, with
$e^{2\mu}\approx|\phi-\phi_{\pm}|^{-1}$. The lower branch $\phi_-$
exists only for $d\ge4$.

\end{description}

In Figures 5, 6 and Figures 7, 8 we plot the curves $U=0$ (dashed
lines) together with the ones for which $d\mu/d\phi=\pm1$ (bold lines)
for cases $d=3$ and $d=6$. Notice the {\em difference} between the
$d<4$ and the $d\ge4$ cases.  For each region we further indicate
whether $\Psi$ is expected to be oscillatory ({\it osc.}) or
exponential ({\it exp.}).

In the following subsections we shall analyse in some detail different
physical situations and derive the corresponding Hartle-Hawking
(no-boundary) wave-function.  We shall employ the transformation
(\ref{eq:4.7}), after which the Wheeler-DeWitt equation takes the
general form 
\beq \left[\frac{\partial^2}{\partial\tilde{\mu}^2}-
\frac{\partial^2}{\partial\tilde{\phi}^2}+U(\tilde{\mu},\tilde{\phi})\right]
\Psi(\tilde{\mu},\tilde{\phi})=0.
\label{eq:WDW}
\eeq We can anticipate that sub-sections {\bf 3.4} and {\bf 3.5}
contain the most interesting physical results as far as the process of
compactification is concerned.

\subsection{Wave function for $\mu>0$ and $\phi\ll0$}

\indent

This case represents the physical situation prior to the
compactification process.  For $\mu> 0$ (i.e. $a > 0$) and $\phi\ll 0$
(i.e. $b \rightarrow 0$ with $U\gg 1$) the potential (\ref{eq:3.3})
becomes 
\beq
U(\mu,\phi)\approx\frac{2k\Lambda}{9\pi}e^{6\mu-(d+4)\alpha\phi}~,
\label{eq:POT}
\eeq
and we can distinguish two situations:
\begin{enumerate}
\item[(a)]$d<4$, for which we can choose 
$\sinh\theta=\frac{6}{\omega}$ and hence $U\approx U(\tilde{\phi})=
\frac{2k\Lambda}{9\pi}e^{-\omega\tilde{\phi}}$ 
\item[(b)] $d\ge4$, for which we can choose 
$\cosh\theta=\frac{6}{\omega}$ and hence $U\approx U(\tilde{\mu})=
\frac{2k\Lambda}{9\pi}e^{\omega\tilde{\mu}}$,
\end{enumerate}
where $\omega^2=|24(-d^2+d+8)/d(d+2)|$, with $\omega>0$.

We can now 
solve eq. (\ref{eq:WDW}) with 
(\ref{eq:POT}) 
by separation of variables to find that for $d<4$,
the solution is a combination of the Bessel functions of the first kind,
$I_{\nu}(z)$, and of the second kind, $K_{\nu}(z)$. For $d\ge4$, we have 
a combination of the modified Bessel functions of the first kind, 
$J_{\nu}(z)$, and of the second kind, $Y_{\nu}(z)$. The study of
the boundary conditions  carried out above  allows us to pick the
appropriate Bessel fuction:
\begin{eqnarray}
\Psi(\tilde{\mu},\tilde{\phi})  \approx  e^{\pm\sqrt\epsilon \tilde{\mu}}
K_{\frac{2}{\omega}\sqrt\epsilon}\left[\frac{2}{\omega}
\left(\frac{2k\Lambda}{9\pi}\right)^{1/2}e^{-\frac{\omega}{2}\tilde{\phi}}
\right] &\mbox{, for $d<4$}~,\label{eq:4.18a}\\
\Psi(\tilde{\mu},\tilde{\phi}) \approx  e^{\sqrt\epsilon \tilde{\phi}}
J_{\frac{2}{\omega}\sqrt\epsilon}\left[\frac{2}{\omega}
\left(\frac{2k\Lambda}{9\pi}\right)^{1/2}e^{\frac{\omega}{2}\tilde{\mu}}
\right] &\mbox{, for $d\ge4$}~,\label{eq:4.18b}\\
\Psi(\tilde{\mu},\tilde{\phi}) \approx
J_{0}\left[\frac{2}{\omega}
\left(\frac{2k\Lambda}{9\pi}\right)^{1/2}e^{\frac{\omega}{2}\tilde{\mu}}
\right] &\mbox{, for $d\ge 19$ and $v_1=0$}~,
\label{eq:4.18c}
\end{eqnarray}
where $e^{\pm\sqrt{\epsilon}\tilde{\mu}}$ means a combination of
$e^{\sqrt{\epsilon}\tilde{\mu}}$ and $e^{-\sqrt{\epsilon}\tilde{\mu}}$,
and
$\epsilon$ is the separation constant, which is determined by matching
this solution onto the solution in the adjacent region (one can also see that
$\epsilon\approx 0$). In (\ref{eq:4.18a})-(\ref{eq:4.18c}) we have assumed that
$\epsilon\ge0$. The case $\epsilon<0$ is not consistent with the
Hartle-Hawking boundary conditions for a wave function of the type
$I_{{\rm const.}\sqrt{\epsilon}}(z)$.
Notice that, as expected, $d<4$ implies an exponential behaviour, while 
$d\ge 4$ corresponds to an oscillatory one.

\subsection{Wave function for $\mu\gg 1$ and $\phi\gg1$} 

\indent 

This case corresponds to the situation where the radii of the $S^3$
and $S^d$ sections are large.  For $\mu\gg 1$ and $\phi\gg1$ one has to deal
with two regions separated by $\mu=\frac{d\alpha}{2}\phi$, on which
different behaviours are expected.  On the lower region (1) in Figure
7 ($\mu<\frac{d\alpha}{2}\phi$) the potential is approximately given
by 
\beq U(\mu,\phi)\approx\left\{\begin{array}{cl} \displaystyle
-e^{4\mu} &\mbox{, for $v_1/v_2=0$}\\ \displaystyle
e^{2\mu+d\alpha\phi}\frac{3\pi}{k}\frac{v_1}{v_2}b_0^2 &\mbox {, for
$v_1/v_2\neq 0$.}
\end{array}
\right.
\label{eq:4.21}
\eeq

For $v_1/v_2\neq 0$ we can choose $\sinh\theta=-\sqrt{(d+2)/2(d-1)}$, so that
$U\approx U(\tilde{\phi})=e^{\bar\omega\tilde{\phi}}
\frac{3\pi}{k}\frac{v_1}{v_2}b_0^2$, where $\bar\omega=\sqrt{8(d-1)/(d+2)}$.
The solution is then 
\beq
\Psi(\tilde{\mu},\tilde{\phi})\approx e^{\sqrt{\bar\epsilon}\tilde{\mu}}
K_{\frac{2}{\bar\omega}\sqrt{\bar\epsilon}}\left[\frac{2}{\bar\omega}
\left(\frac{3\pi}{k}\frac{v_1}{v_2}b_0^2\right)^{1/2}
e^{\frac{\bar\omega}{2}\tilde{\phi}}
\right]~,
\label{eq:4.22}
\eeq
where $\bar\epsilon$ is the separation constant.

For $v_1/v_2=0$ the wave function is a combination of 
$K_0(z)$ and $I_0(z)$, with 
$z=\frac{1}{2}e^{2\mu}$. These solutions are, as expected, exponential type 
and are also valid in the region $\phi\gg1$ and $\mu<0$.

For the other case (region (2) in Figure 7), 
we have $U\approx e^{6\mu-d\alpha\phi}
\frac{2k\Lambda}{9\pi}$. Choosing $\sinh\theta=\sqrt{d/2(d+3)}$ we get
$U\approx U(\tilde{\mu})=e^{\tilde\omega\tilde\mu}\frac{2k\Lambda}{9\pi}$,
with $\tilde\omega=\sqrt{24(d+3)/(d+2)}$, and $\Psi$ is a combination of
$J_{\nu}(z)$ and $Y_{\nu}(z)$, with $\nu=\frac{2}{\tilde\omega}
\sqrt{\tilde\epsilon}$ and $z=\frac{2}{\tilde\omega}\left(
\frac{2k\Lambda}{9\pi}\right)^{1/2}e^{\frac{\tilde\omega}{2}\tilde\mu}$, 
$\tilde\epsilon$ being a separation constant.
This solution is, as expected, oscillatory.

\subsection{Wave function for $\mu\ll0$}

\indent

This case corresponds to a 4-dimensional physical Universe  
at a very early stage and with a generic $S^d$ section.
In the region $\mu\ll0$ (i.e., $a(t) \rightarrow 0$)
and $\phi>0$ the potential is also given by 
(\ref{eq:4.21}). For $v_1/v_2\neq 0$ we obtain 
\beq
\Psi\approx e^{\pm\sqrt{\hat\epsilon}\tilde\mu}
I_{\frac{2}{\hat\omega}\sqrt{\hat\epsilon}}\left[\frac{2}{\hat\omega}
\left(\frac{3\pi}{k}\frac{v_1}{v_2}b_0^2\right)^{1/2}
e^{\frac{\hat\omega}{2}\tilde{\phi}}
\right]~,\label{eq:4.23}
\eeq
while for $v_1/v_2=0$ we have $\Psi(\mu,\phi)\approx I_0\left[
\frac{1}{2}e^{2\mu}\right]$. In both cases the behaviour is exponential.
These solutions also apply for $\phi<0$ and $\mu<\frac{\alpha(d+4)}{2}\phi$.

For the paprticular situation 
where  $\mu\ll0$ together with  $\phi\ll0$,  we further 
distinguish two different situations:
\begin{enumerate}
\item[(a)] For $\mu>\frac{(d+2)\alpha}{2}\phi$ we expect a behaviour 
similar to the one found for $\mu>0$ and $\phi\ll 0$ (see subsection 3.1).
\item[(b)] As for the region $\frac{(d+2)\alpha}{2}\phi<\mu<\frac{(d+4)
\alpha}{2}\phi$, this is a transition region and one should expect a 
mixture of the previous wave functions.
\end{enumerate}

\subsection{Wave function in the neighbourhood of $\phi=\phi_{\rm max}$}

\indent 

We shall now obtain approximate solutions in the neighbourhood of 
$\phi=\phi_{\rm max}$  using the semiclassical approximation to the path 
integral (\ref{eq:3.4})
\beq
\Psi(\mu,\phi)\approx A(\mu,\phi)e^{-S_{\rm E}(\mu,\phi)},
\label{eq:4.25}
\eeq
where $\phi_{\rm max}$ is the local maximum of $U(\mu={\rm constant},\phi)$ 
and is  
given approximately by $e^{-2\alpha\phi_{\rm max}}=d/d+4$. 
This corresponds to the physical state of our universe where 
the extra $d$-dimensional space is at an equilibrium point, 
corresponding to its maximum value.

Using then the classical
field equations of motion obtained from $S_{\rm eff}$ to
integrate the Euclidean action we get for $v_1/v_2=0$: 
\beq
S_{\rm E}=\frac{3}{16k^2\Omega}\left[
\left(1-\left(\frac{4k}{3}\right)^2e^{2\mu}\Omega\right)^{3/2}-1\right],
\eeq
where the potential $\Omega (\mu, \phi)$, given by
\beq
\Omega (\mu, \phi) = 
e^{-d\alpha\phi} \frac{\Lambda}{8\pi k} \left(e^{-2\alpha\phi} - 1\right)^2
+     
e^{d\alpha\phi -4 \mu} \left(\frac{27\pi b_0^2}{16 k^3}
\frac{v_1}{v_2 }\right)~,
\label{eq:3.18c}
\eeq
was assumed to be approximately 
constant near $\phi=\phi_{\rm max}$. Hence, in the region $U<0$:
\beq
\Psi\approx A(\mu,\phi)\exp\left[\frac{3}{16k^2\Omega}
\right]\exp\left[-\frac{3}{16k^2\Omega}
\left(1-\left(\frac{4k}{3}\right)^2e^{2\mu}\Omega\right)^{3/2}\right],
\label{eq:4.28}
\eeq
where the prefactor $A$ is such that it verifies the condition 
$A(-\infty,\phi)=1$.

In the region $U>0$ the wave function becomes oscillatory, and the WKB
procedure shows that
\beq
\Psi\approx B(\mu,\phi)\exp\left[\frac{3}{16k^2\Omega}
\right]\cos\left[\frac{3}{16k^2\Omega}
\left(\left(\frac{4k}{3}\right)^2e^{2\mu}\Omega-1\right)^{3/2}-
\frac{\pi}{4}\right]~.
\label{eq:4.29}
\eeq
Replacing (\ref{eq:4.29}) in the Wheeler-DeWitt equation one obtains the 
prefactor
\beq
B(\mu,\phi)\approx e^{-\mu}\left[\left(\frac{4k}{3}\right)^2e^{2\mu}\Omega-1
\right]^{-1/4}~.
\eeq

For $v_1/v_2\neq 0$ these results are still valid for $\mu>0$. For
$\mu<0$
we expect the behaviour described in subsection 3.3.

\subsection{Wave function in the neighbourhood of $\phi=0$ and large $\mu$}

\indent

Finally, we consider the case where the 4-dimensional physical Universe 
is in a stage of large $S^3$ radius and with $b \sim b_0$.
In the neighbourhood of $\phi=0$, at the minimum of 
$U(\mu={\rm constant},\phi)$,
we consider the dominant term of the potential for large $\mu$:
\beq
U(\mu,\phi)\approx e^{6\mu-d\alpha\phi}\frac{2k\Lambda}{9\pi}
\left(e^{-2\alpha\phi}-1\right)^2\approx 
e^{6\mu}\phi^2 \frac{8\alpha^2 k\Lambda}{9\pi}~.
\label{eq:3.19}
\eeq
Notice that the potential vanishes for $\phi=0$ and that in (\ref{eq:3.19})
we exhibit the dominant term for values of $\phi$
around the minimum. Quadratic potentials of this kind
are found in  massive scalar field models \cite{swh}.

We now perform a simple change of variables
\beq
\begin{array}{l}
x=e^{3\mu}\\
y=e^{-2\alpha\phi}
\end{array}
\eeq
from which yields the Wheeler-DeWitt equation:
\beq
\left[9x^2\frac{\partial^2}{\partial x^2}-4\alpha^2y^2
\frac{\partial^2}{\partial y^2}+9x\frac{\partial}{\partial x}-4\alpha^2
y\frac{\partial}{\partial y}+x^2y^{d/2}(y-1)^2\frac{2k\Lambda}{9\pi}
\right]\Psi(x,y)=0~.
\eeq

As we are interested in the limit $x\ll1$ and $y\approx1$, we actually have to
solve:
\beq
\left[x^2\frac{\partial^2}{\partial x^2}+x\frac{\partial}{\partial x}
+\frac{1}{9}x^2y^{d/2}(y-1)^2\frac{2k\Lambda}{9\pi}
\right]\Psi(x,y)=0~.
\eeq

Thus, choosing $z=\frac{1}{3}\sqrt{\frac{2k\Lambda}{9\pi}}xy^{d/4}|y-1|$, 
one easily
sees that $\Psi$ is a combination of Bessel functions $J_0(z)$ and $Y_0(z)$,
where $z=\sqrt{\frac{2k\Lambda}{9\pi}\frac{2\alpha}{3}}e^{3\mu}|\phi|$.
If $\Psi\propto J_0(z)$ then, as $z\rightarrow 0$, the wave function 
behaves as $\Psi\approx 1-z^2/4$.
If, on the other hand, $\Psi$ also depends on $Y_0(z)$, then, as 
$z\rightarrow 0$, $\Psi$ behaves asymptotically as $\Psi\approx \frac{2}{\pi}
\ln\frac{z}{2}$. This behaviour is depicted in Figure 9.

According to the standard interpretational rules of quantum cosmology
(see for instance Ref. \cite{vil}), the probabilistic interpretation
of the wave function does make sense in the classical and the
semiclassical regions.  Therefore, as the large $\mu$ region
corresponds to a classical region, the fact that the wave function is
highly peaked around $\phi=0$ means that the most probable
configuration does indeed correspond to solutions with
compactification for expanding external spacetime.  In the next
section we shall draw additional physical information concerning some
of the solutions in this section.

\section{Interpretation of the wave function}

\indent

In order to interpret the wave function we shall use the trace of the square 
of the extrinsic curvature, $K^2=K_{\hat I\hat J}K^{\hat I\hat J}$, 
to see whether the wave function in the semiclassical limit corresponds to a 
Lorentzian or to a Euclidean geometry. This is justified as 
the Wheeler-DeWitt equation is the same from whatever metric (Lorentzian or 
Euclidean) one derives it.
The extrinsic curvature is a measure of the variation of the normal 
to the hypersurfaces of constant time, and is given by:
\beq
K_{\hat I\hat J}=N^{-1}\left(-\frac{1}{2}\frac{\partial h_{\hat I\hat J}}
{\partial t}+\nabla_{\hat J}N_{\hat I}\right)~,
\eeq
where $h_{\hat I\hat J}$ is the $d+3$-metrics and 
$N_{\hat I}$ are the components of the shift-vector.
>From (\ref{eq:2.11}) and using 
(\ref{eq:1.10}) we obtain
\beq
K^2=-e^{-6\mu+d\alpha\phi}\frac{3\pi}{2k}\left[9\frac{\partial^2}
{\partial\mu^2}+\left(\frac{d\alpha}{2}\right)^2
\frac{\partial^2}{\partial\phi^2}+
3d\alpha\frac{\partial^2}{\partial\mu\partial\phi}\right]~.
\eeq
Performing the Lorentz-type transformation (\ref{eq:4.7}) with $\sinh\theta=
\sqrt{d/2(d+3)}$, and using $\tilde\omega=\sqrt{24(d+3)/(d+2)}$, 
$K^2$ simplifies to
\beq
K^2=-e^{-\tilde\omega\tilde\mu}\frac{9\pi}{k}\left(\frac{d+3}{d+2}\right)
\frac{\partial^2}{\partial\tilde\mu^2}~,
\eeq
and we see that, in regions where the wave function behaves as an exponential 
the quantity $K^2\Psi/\Psi$ is negative. Therefore, in the classical limit, 
$K$ is imaginary 
and we have a Euclidean ($d+4$)-geometry. When the wave function is 
oscillatory,
the corresponding $K$ is real, and the ($d+4$)-geometry is Lorentzian.
Note that a Lorentz geometry corresponds to a classical state of the 
Universe, while a Euclidean one is normally associated to a quantum or 
tunneling state.
As shown in Figures 7 and 8 there exist, for $d\ge 4$, well defined 
Lorentzian regions for {\em different} values of the ratio $v_1/v_2$.
These regions are however, inexistent when $d<4$ as depicted in Figures
5 and 6.

In the oscillatory region, the wave function can be further interpreted 
using the WKB approximation $\Psi={\rm Re}\left(Ce^{iS}\right)$, where 
$S$ is a rapidly varying phase and $C$ a slowly varying prefactor. One 
chooses $S$ to satisfy the classical Hamilton-Jacobi equation
\beq
-\left(\frac{\partial S}{\partial\mu}\right)^2+\left(\frac{\partial S}
{\partial\phi}\right)^2+U(\mu,\phi)=0~.
\label{eq:5.8}
\eeq
The significance of $S$ becomes evident when operating $\pi_{\mu}$ on $\Psi$
(for $\pi_{\phi}$ the procedure is analogous):
\beq
\pi_{\mu}\Psi=\left[\frac{\partial S}{\partial\mu}-i\frac{\partial}
{\partial\mu}\ln C\right]\Psi~.
\eeq
Since in the WKB approximation we assume $|\frac{\partial S}{\partial\mu}|\gg
|\frac{\partial}{\partial\mu}\ln C|$, we have
\beq
\pi_{\mu}=\frac{\partial S}{\partial\mu}~,~
\pi_{\phi}=\frac{\partial S}{\partial\phi}~.
\label{eq:5.11}
\eeq

The wave function corresponds then to a two-parameter subset of 
solutions which obey (\ref{eq:5.11}) and that can be regarded as providing 
the boundary conditions for the classical solutions.
We shall now try to obtain an approximate solution for the Hamilton-Jacobi 
equation (\ref{eq:5.8}) in the region close to the space segment $U=0$
and $\phi=\phi_{\rm max}$ as it is there that classical trajectories start.
Assuming that $S$ is separable and that $|\frac{\partial S}{\partial\mu}|\gg
|\frac{\partial S}{\partial\phi}|$ we can use a series expansion around 
$\phi=\phi_{\rm max}$ to obtain
\beq
S\approx\pm\frac{e^{3\mu}}{3}\left(\frac{2k\Lambda}{9\pi}\right)^{1/2}
\left(\frac{d}{d+4}\right)^{d/4}\left[\frac{4}{d+4}-E\left(e^{-2\alpha\phi}-
\frac{d}{d+4}\right)^2\right]~,
\label{eq:5.13}
\eeq
where $E=\frac{3}{16}\frac{(d+2)(d+4)}{d}\left[\left(
1+\frac{4}{3} \left(\frac{d+4}{d+2}\right)\right)^{1/2}-1\right]$.
The upper (lower) sign on (\ref{eq:5.13}) corresponds to a collapsing
(expanding) Universe. This result agrees with (\ref{eq:4.29}). Using
(\ref{eq:5.11}) and (\ref{eq:1.10}) we have, for the gauge $N=1$,
\begin{eqnarray}
\dot\mu & \approx & \mp\left(\frac{\Lambda}{3}\right)^{1/2}
\left(\frac{d}{d+4}\right)^{d/4}\left[\frac{4}{d+4}-E\left(e^{-2\alpha\phi}-
\frac{d}{d+4}\right)^2\right]~,\label{eq:5.14}\\
\dot\phi & \approx & \pm\left(\frac{\Lambda}{3}\right)^{1/2}
\frac{4\alpha}{3}Ee^{-2\alpha\phi}\left(e^{-2\alpha\phi}-
\frac{d}{d+4}\right) \label{eq:5.15}~.
\end{eqnarray}

If $\phi_0$, the initial value of $\phi$, is close to $\phi_{\rm max}$, then
$\dot\phi$ will be very small and the scale factor $a(t)$ will grow
exponentially like $\exp\left[\left(\frac{\Lambda}{3}\right)^{1/2}
\left(\frac{d}{d+4}\right)^{d/4}\frac{4}{d+4}t\right]$ for an expanding
Universe. Given the fine-tuning (\ref{eq:2.2}) this last expression
becomes
\beq
a(t)\approx\exp\left[\frac{1}{b_0}
\left(\frac{d}{d+4}\right)^{(d+2)/4}
\left(\frac{d-1}{3(d+4)}\right)^{1/2}t\right]~,
\label{eq:5.16}
\eeq
which gives $a(t)\approx\exp\left(\frac{1}{b_0}\frac{1}{\sqrt3 e}t\right)$
for $d\rightarrow +\infty$.

Thus, we confirm the expectation that $\phi$ configurations to which the 
main contribution to the potential after compactification is an effective 
cosmological constant, do correspond, in the semiclassical regime, 
to inflationary solutions for expanding universes.

\subsection{Wave function for the vacuum configuration $v_2=V_2({\bf g}^v$)}

\indent

Throughout the previous sections we have assumed that $v_2>0$. 
This corresponds to the choice $\bf{g}=\bf{0}$ for 
the potential (\ref{eq:2.20b}), 
which is obviously associated  to a classically unstable situation. 
Nevertheless, since the wave function can be interpreted, at least in 
a semiclassical situation, as giving the probability
of a certain configuration, one expects, for consistency, to have
the wave function peaked around $\bf{g}=\bf{0}$ when unfreezing $v_2$ and 
varying $\bf g$. This means that the most probable configuration 
should  correspond to the choice $\bf{g}=\bf{0}$.

The dependence of $\Psi$ on $v_2$ can be seen fixing the value of the
gauge coupling constant, $e$, and rewriting the term
$\frac{2k\Lambda}{9\pi}$ in potential (\ref{eq:3.3}) as
$\left(\frac{e}{12\pi}\right)^2\frac{d(d-1)}{2 v_2}$ (where
(\ref{eq:2.2}) was used). Furthermore, using the value of the radius
of compactification, $b_0^2=\frac{16\pi kv_2}{e^2}$, we can see that
the term $b_0^2\frac{v_1}{v_2}$=$\frac{v_1}{16\pi k}e^2$ in
(\ref{eq:3.3}) does not depend on $v_2$. We hence conclude that
solutions depending on $\Lambda$ will depend on $v_2^{-1}$.

We are only interested in regions where $\mu>0$ ($a>0$), i.e. in
regions where the probabilistic interpretation can be unambiguously
used, from which implies that we have the following cases:
\begin{enumerate}
\item[(a)] For $\mu>0$ and $\phi\ll0$ (i.e, $b(t) \rightarrow 0$),
$\Psi\propto K_{\nu} (\Lambda^{1/2})$ (Figure 10) or $\Psi\propto
J_{\nu}(\Lambda^{1/2})$ (Figure 11) according to the value of $d$
(cf. wave functions (53) and (54)).
\item[(b)] For $\mu\gg 1$, $\phi\gg 1$ and
$\mu>\frac{d\alpha}{2}\phi$, $\Psi$ is a combination of
$J_{\nu}(\Lambda^{1/2})$ and $Y_{\nu}(\Lambda^{1/2})$.  If ${\bf g}$
is not too large the behavior of $Y_{\nu}$ is similar to the one of
$J_{\nu}$ (Figure 11).
\item[(c)]
For $\phi\approx \phi_{\rm max}$ and $\mu>0$ the wave function is given by 
either (\ref{eq:4.28}) or (\ref{eq:4.29}) (Figure 12).
\item[(d)] Finally, for $\phi\approx 0$ , the wave function, $\Psi$,
is a combination of $J_0(\Lambda^{1/2})$ and $Y_0(\Lambda^{1/2})$,
whose behavior is similar to the one depicted in Figure 11.
\end{enumerate}

Notice that when $\Psi$ is oscillatory, the peak for $\bf{g}=\bf{0}$
will disappear for certain values of $\mu$.  Nevertheless, we have
always $\Psi(|{\bf g}|=\pm 1)=0$.  We can therefore conclude that we
do observe the expected maximum of the wave function for
$\bf{g}=\bf{0}$.

\section{Conclusions}

\indent 

In this paper we have obtained solutions of the Wheeler-DeWitt
equation derived from the effective model that arises from
dimensionally reducing to one dimension the Einstein-Yang-Mills
generalized Kaluza-Klein theory in $D=4+d$ dimensions. We considered a
${\bf R} \times S^3 \times S^d$ topology and the corresponding
Hartle-Hawking boundary conditions.  The dimensional reduction was
achieved by restricting the field configurations to be homogeneous and
isotropic through coset space compactification as indicated in
Sections 1 and 2.  This model of compactification has been proposed in
Refs. \cite{krt, bkm}.  In particular, the crucial role played by the
external space components of the gauge field in order to achieve
classically as well as semiclassically stable compactifications was
shown in Ref. \cite{bkm}.

In Section 2 we have presented the
most salient features of the model and set up the Hamiltonian
constraint which allows us to obtain the Wheeler-DeWitt equation to
study the compactification process from the quantum mechanical point
of view. Notice that in our model the gauge fields associated angular
momentum is also constrained to vanish.  The richness of our effective
model (\ref{eq:2.17}) is quite evident.  In this reduced model the
gauge field has {\em non-vanishing} time-dependent components in {\em
both} the external and internal spaces.  Moreover, we have also two
time-dependent scalar fields, the dilaton and the inflaton. This
contrastes with previous work in the literature, where either static
magnetic monopole configurations with non-zero components only in
$I^d$ or scalar fields were present.

In section 3 we have obtained no-boundary solutions of the
Wheeler-DeWitt equation which exhibit very interesting features.  The
term $e^{d\alpha\phi -4\mu} \left(\frac{6\pi}{k}\right)^2
\frac{3}{4e^2}v_1$ in (\ref{eq:1.18c}) establishes that the external
spatial dimensions and the internal $d$-dimensions are at the {\em
same footing} in the early Universe prior to compactification, i.e.
when $\mu \ll 0$. It is only through the expansion of the external
dimensions (increase of $\mu$) that compactification ($b \rightarrow
b_0$) is achieved. Thus, it is the dynamics of the $3-$dimensional
physical space which induces the evolution of $I^d$ towards
compactification.

We also find that stable compactifying solutions do correspond to
extrema of the wave function of the Universe showing that the process
of compactification does indeed takes place for expanding
universes. Furthermore, our analysis indicates that the main
properties of the Hartle-Hawking wave function do depend on the
following two features. On the one hand, on a non-vanishing
contribution to the potential (\ref{eq:3.3}) of the external physical
space dimensions of the gauge field, a feature already found in the
classical analysis of Ref. \cite{bkm}. On the other hand, also on the
number, $d$, of internal space dimensions. In the case we set the
contribution of the external space dimensions of the gauge field to
the potential (\ref{eq:3.3}) to vanish, we find that we recover the
main aspects of the discussion of Ref. \cite{jh}, where
compactification was discussed in the framework of an Einstein-Maxwell
model with a magnetic monopole configuration whose gauge (Maxwell)
field contribution was non-vanishing only for the internal space. The
same can be said about Ref. \cite{cw}, where a stable compactification
was achieved through the non-vanishing contribution of the internal
components of a ($D-4$)th rank antisymmetric tensor field. Finally, we
also find that for expanding models, inflationary solutions can be
predicted, as shown in section 4, if in the semiclassical regime the
potential is essentially given by an effective cosmological constant.
   
\vspace{1cm}

{\large\bf Acknowledgements} 

\noindent
One of us (P.V.M.) gratefully acknowledges the support of the
JNICT/PRAXIS XXI Fellowship BPD/6095/95 The authors are grateuful to
A. Zhuk, V. Ivashchuck, V. Melnikov, M. Rainer, K. Bronnikov for
conversations and discussions which have motivated this work and to
Yu.A. Kubyshin for valuable discussions and suggestions.

\newpage

{\Large \bf Appendix}

\vskip 0.3cm
{\Large \bf \indent Hartle-Hawking proposal 
and its generalization to higher spacetime dimensions }

For clarification purposes, let us briefly outline here the main
features of the Hartle-Hawking proposal \cite{hh} and its
generalization to higher spacetime dimensions \cite{jh} (see also
ref. \cite{cw,chinaprd,stong,chie}).  In quantum cosmology it is
assumed that the quantum state of a D=4 Universe is described by a
wave function $\Psi[h_{ij}, \Phi]$, which is a functional of the
spatial 3-metric, $h_{ij}$, and matter fields generically denoted by
$\Phi$ on a compact 3-dimensional hypersurface $\Sigma$. The
hypersurface $\Sigma$ is then regarded as the boundary of a compact
4-manifold ${\cal M}^4$ on which the 4-metric $g_{\mu\nu}$ and the
matter fields $\Phi$ are regular. The metric $g_{\mu\nu}$ and the
fields $\Phi$ coincide with $h_{ij}$ and $\Phi_0$ on $\Sigma$ and the
wave function is then defined through the path integral over
4-metrics, $^{4}g$, and matter fields:
\begin{equation}
\Psi[h_{ij}, \Phi_0] = \int_{{\cal C}} D[^{4}g] D[\Phi]
\exp\left(-S_{E}[^{4}g, \Phi]\right) ~,
\label{eq:1.1}
\end{equation}
where $S_E$ is the Euclidean action and $\cal C$ is the class of
4-metrics $g_{\mu\nu}$ and regular fields $\Phi$ defined on Euclidean
compact manifolds $M^4$ and with {\em no other} boundary than
$\Sigma$.  An extension of the Hartle-Hawking proposal for universes
with $D > 4$ dimensions was first discussed in Ref. \cite{jh}.  Let us
summarize it, mentioning some of its difficulties and comparing it
with the $D=4$ case.

In $D=4$ the theory of cobordism \cite{stong} guarantees that for all
compact 3-surfaces there {\em always} exists a compact 4-dimensional
manifold such that $S^3$ is the {\em only} boundary, or equivalently,
all 3-dimensional {\em compact} hypersurfaces are cobordant to zero
\cite{stong}.  Let us now consider the case for $D>4$. In these
$D$-dimensional models, the wave function would be a functional of the
$(D-1)$ spatial metric, $h_{IJ}$, and matter fields, $\Phi$, on a
($D-1$)-hypersurface, $\Sigma_{D-1}$ and is defined as the result of
performing a path integral over all compact $D$-metrics and regular
matter fields on $M^{D}$, that match $h_{IJ}$ and the matter fields on
$\Sigma_{D-1}$.

Let us then start by assuming that the $(D-1)$-surface $\Sigma_{D-1}$
does not possess any disconnected parts \cite{jh}. Would there always
be a $D$-dimensional manifold ${\cal M}^D$ such that $\Sigma_{D-1}$ is
the only boundary ?  In higher dimensional manifolds however, this is
not guaranteed. There exist compact $(D-1)$-hypersurfaces
$\Sigma_{D-1}$ for which there is no compact D-dimensional manifold
such that $\Sigma_{D-1}$ is the only boundary.  This seems to indicate
that in $D > 4$ dimensions there are configurations which cannot be
attained by the sum over histories in the path integral.  The wave
function for such configurations would therefore be zero.  In
ref. \cite{jh} this situation was be circunvented so as to obtain
non-zero wave-functions for such configurations, namely by dropping
the assumption that the $(D-1)$-surface $\Sigma_{D-1}$ does not
possess any disconnected parts.

As described in \cite{jh}, if one assumes that the hypersurfaces
$\Sigma_{D-1}$ consist of any number $n>1$ of disconnected parts
$\Sigma_{D-1}^{(n)}$, then one finds that the path integral for this
disconnected configuration involves terms of two types. The first type
consists of disconnected $D-$manifolds, each disconnected part of
which closes off the $\Sigma^{(n)}_{D-1}$ surfaces separately. These
will exist only if each of the $\Sigma^{(n)}_{D-1}$ are cobordant to
zero, but this may not always be the case. There will indeed be a
second type of term which consists of connected $D-$manifolds which
just plainly joins some of the $\Sigma_{D-1}^{(n)}$ together.  This
second type of manifold will {\em always} exist in {\em any} number of
dimensions, providing the $\Sigma_{D-1}^{(n)}$ are similar
topologically, i.e. have the same {\em characteristic numbers}
\cite{stong}. The wave function of any $\Sigma_{D-1}^{(1)}$ surface
which is not cobordant to zero would be different from zero and
obtained by assuming the existence of other surfaces of suitable
topology and then summing over all compact $D-$manifolds which join
these surfaces together.  Thus, given a compact $(D-1)$ hypersurface
$\Sigma_{D-1}$ which is not cobordant to zero, a non-zero amplitude
could be obtained by {\em assuming} it possesses disconnected parts.

However, the above considerations for disconnected pieces and generic
$\Sigma_{D-1}$ surfaces would spoil the Hartle-Hawking prescription
since the manifold would have more than one boundary.  In other words,
the general extension above discussed would imply a description in
terms of propagation between such generic $\Sigma_{D-1}$ surfaces.
The wave function would then depend on every piece and not on a single
one as advocated in \cite{jh}.  Nevertheless, if we restrict
ourselves, as we do in the present paper, to the case of a truncated
model with a global topology given by a product of a 3-dimensional
manifold to a $d$-dimensional one, then the spacelike sections always
form a boundary of a $D$-dimensional manifold with no other boundaries
\cite{chie}. Since hypersurfaces $S^3 \times S^d$ are always cobordant
to zero, it implies that for spacetimes with topology ${\bf R} \times
S^3 \times S^d$ the Hartle-Hawking proposal can be always implemented,
and thus we can consider the original no-boundary proposal in our
study.

\newpage

{\bf Figure Captions}

\vskip 0.4cm
{\bf Figure 1} \\
 Potential $U(\mu={\rm constant},\phi)$ for some 
values of $\Lambda$ and $d=6$ ((a) $\Lambda>\frac{c_2}{16\pi k}$, 
(b) $\frac{c_1}{16\pi k}<\Lambda<\frac{c_2}{16\pi k}$, 
(c) $\Lambda<\frac{c_1}{16\pi k}$). 

\vskip 0.4cm
{\bf Figure 2} \\
 Potential $U(\mu,\phi)$ for $d=6$ and large $\mu$ ($\mu>\mu_{\rm c}$, see
Figure 3).

\vskip 0.4cm
{\bf Figure 3} \\
 Potential $\Omega(\mu={\rm constant}, \phi)$ for $d=6$ and some 
values of $\mu$.

\vskip 0.4cm
{\bf Figure 4} \\
$U=0$ curves in the $\mu \phi$-plane for $d=6$ and different 
values of the ratio $\frac{v_1}{v_2}$ ((a) $\frac{v_1}{v_2}=0$, 
\break (b) $\frac{v_1}{v_2}=\frac{1}{3}$, (c) $\frac{v_1}{v_2}=1$). 
  
\vskip 0.4cm
{\bf Figure 5} \\
$U=0$ (dashed) and null curves (bold) in the $\mu \phi$-plane for $d=3$ and 
$\frac{v_1}{v_2}=0$.

\vskip 0.4cm
{\bf Figure 6} \\
$U=0$ (dashed) and null curves (bold) in the $\mu \phi$-plane for $d=3$ and 
$\frac{v_1}{v_2}=1$.

\vskip 0.4cm
{\bf Figure 7} \\
$U=0$ (dashed) and null curves (bold) in the $\mu \phi$-plane for $d=6$ and 
$\frac{v_1}{v_2}=0$.

\vskip 0.4cm
{\bf Figure 8} \\
$U=0$ (dashed) and null curves (bold) in the $\mu \phi$-plane for $d=6$ and 
$\frac{v_1}{v_2}=1$.

\vskip 0.4cm
{\bf Figure 9} \\
Wave function in the neighbourhood of $\phi=0$.

\vskip 0.4cm
{\bf Figure 10} \\
Module of the wave function for $\mu > 0$ and $\phi\ll0$.

\vskip 0.4cm
{\bf Figure 11} \\
Module of the wave function in the region (2) of Figure 7.

\vskip 0.4cm
{\bf Figure 12} \\
Module of the wave function in the neighbourhood of 
$\phi=\phi_{\rm max}$ and $\mu >0$.


\end{document}